\documentclass{vgtc}                          

\ifpdf
  \pdfoutput=1\relax                   
  \usepackage{graphicx}                
  \DeclareGraphicsExtensions{.pdf,.png,.jpg,.jpeg} 
\else
  \usepackage{graphicx}                
  \DeclareGraphicsExtensions{.eps}     
\fi%

\graphicspath{{figures/}{pictures/}{images/}{./}} 

\usepackage{microtype}                 
\usepackage{tabularx}
\PassOptionsToPackage{warn}{textcomp}  
\usepackage{textcomp}                  
\usepackage{mathptmx}                  
\usepackage{times}                     
\usepackage{cite}                      
\usepackage{tabu}                      
\usepackage{booktabs}                  
\usepackage{enumitem}

\usepackage{xcolor}
\definecolor{steelblue}{RGB}{70, 130, 180}

\newcommand{\parahead}[1]
{%
  \vspace{0.07in}%
  \noindent%
  \textbf{\textit{#1.}}%
}

\usepackage{soul}
\newcommand{\asLink}[2]{\textcolor{steelblue}{\href{#1}{\ul{#2}}}}

\onlineid{0}

\vgtccategory{Research}

\vgtcinsertpkg

\title{Teaching Critical Visualization: A Field Report}

\author{
Andrew McNutt\thanks{e-mail: andrew.mcnutt@utah.edu}\\
     \scriptsize University of Utah%
\and Shiyi He\\\scriptsize University of Utah
\and Sujit Kumar Kamaraj\\\scriptsize University of Utah
\and Purbid Bambroo\\\scriptsize University of Utah
\and Nastaran Jadidi\\\scriptsize University of Utah
\and John Bovard\\\scriptsize University of Utah
\and Chang Han\\\scriptsize University of Utah
} %

\abstract{
Critical Visualization is gaining popularity and academic focus, yet relatively few academic courses have been offered to support students in this complex area. This experience report describes a recent experimental course on the topic, exploring both what the topic could be as well as an experimental content structure (namely as scavenger hunt). Generally the course was successful, achieving the learning objectives of developing critical thinking skills, improving communication about complex ideas, and developing a knowledge about theories in the area.  While improvements can be made, we hope that humanistic notions of criticality are embraced more deeply in visualization pedagogy. 
} 

\begin{document}


\firstsection{Introduction}

\maketitle

Visualization may finally be having its critical moment. Around a decade ago, Dörk and colleagues proposed Critical InfoVis~\cite{dork2013critical}, which urged consideration of the politics in visualizations, pushing us to consider who, how, and what is being visualized or datafied. While not the first work to consider these ideas, this offers a nice named conceptual starting point. In the years since then, a wide range of explorations and works broadly within this area have been met with broad success. D'ignazio and Klein's Data Feminism~\cite{d2023data} is widely cited and read.
In summer 2025, CG\&A is publishing a two-part issue on critical visualization~\cite{panagiotidou2025critical}.

So, if now is to be the critical moment\footnote{Academic visualization often lags decades behind cartography. To wit, critical cartography has been active at least since the 90s~\cite{crampton2005introduction}.
} for critical visualization, then it is worth pausing to ask: how do we disseminate this somewhat complicated and scholarly topic? While the currently trodden routes of public-facing talks and academic papers are useful, engaging with criticality as part of a visualization education seems valuable.
There have been courses on critical cartography, critical data studies, and post-colonial visualization\cite{dork22Decolonizing}.
So why not critical visualization? Yet it is not clear how to approach such a task, as critical visualization splays across many disciplines, ranging from critical theory to feminism to social technology studies and many topics between.

In the spring semester of 2025, I\footnote{Using the first person here to center the decision-making authority I, the first author, exerted in the course design. The other authors are the students in the class, who are included here to value their efforts.
} taught a graduate computer science course on critical visualization at the University of Utah, CS6967: Critical VIS+HCI.
Enrollment for this 15.5-week course was small (capped at 20 but eventually yielding 6 MS or PhD Students) and kept intentionally away from undergrads,
so that a variety of experiments might be tried.
This article is a field report from those trials, and not, I emphasize, an academic review of criticality in visualization pedagogy (although such work would be valuable).

The course involved two experiments. First, and most saliently, was the teaching of critical visualization. Second, and more experimentally, is the actual structure of the course, which was ``course as scavenger hunt''. At the beginning of the course, a series of activities were suggested with a certain collection of associated point values. For instance, one might submit a zine for 6 points or a project for 10 points. If a student wishes to get an A, then they need only assemble an appropriate collection of submissions towards a maximum of 20. As these submissions are by student design, the only deadline is then the very end of the course.
The syllabus and course content can be found at \asLink{https://www.mcnutt.in/critical-hci-vis-class/}{mcnutt.in/critical-hci-vis-class/} (archived at \asLink{https://osf.io/hxtq2/}{osf.io/hxtq2}).

\section{Reflections}

Broadly, this class felt successful to me as an educator: the learning goals of critical thinking, communication, and theory were met, and students were able to express cogent syntheses of their learning relating to those areas.
For instance, one student explored how the notion of vanguardism might be applied to data feminism. Broadly, I think that this style of course could be useful for more students (I know I wish I'd had something like this in my graduate education). Here I reflect on both of the experiments in the course.

\subsection{Content}

Critical visualization means different things to different people, and so any slice through this vast topic will inevitably leave some things out.
I mainly focused on contemporary feminist and critical visualization, supplemented  by foundational readings in the underpinning theories.
This led us to read works focused on visualization specifically (such as Kennedy et al.'s~\cite{kennedy2016work} explorations of visualization conventions), as well as some original sources on critical topics---such as Horkheimer's~\cite{horkheimer1972traditional} essay debuting critical theory or Haraway's situated knowledges~\cite{haraway2013situated}.

I picked these works because they reflected my understanding (and others' treatments~\cite{hall2022critical, dork2013critical}) of critical visualization; although other paths are possible---for instance, centering post-colonial and post-growth computing offer intriguing conceptual starting points.
The initial corpus from which I refined the course began as a crowd-built list of critical visualization papers on the ``Vis Researchers'' Slack community.
This group conceptualization of critical visualization strongly shaped the resulting course material---those papers not read but listed in that thread are present in the course materials.

Within this setting, I used visualization as the subject of study, rather than the means of study. A traditional visualization class might value data thinking and use the topics covered in example data (e.g., penguins) to support those goals. In contrast, I use visualization (as artifacts) to develop and support criticality.

\begin{figure}
  \centering
  \includegraphics[width=\linewidth]{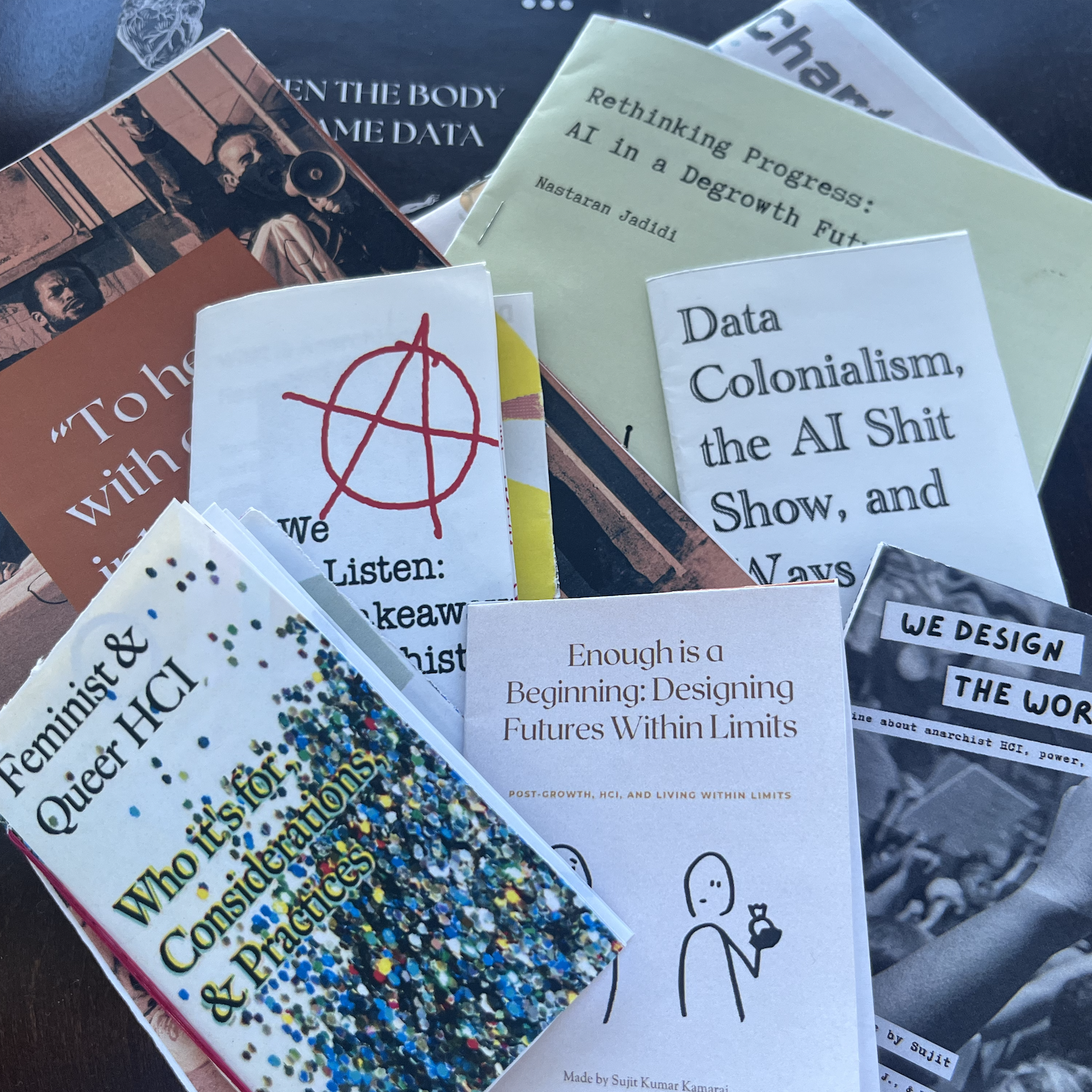}
  \caption{Some students submitted zines to the course. While this is likely a reflection of my own enthusiasm for the form~\cite{mcnutt2021potential}, students seemed to express excitement for the new medium.}
  \label{fig:enter-label}
  \vspace{-2em}
\end{figure}

\parahead{Utility}
In describing this class to colleagues, a frequent question was ``were CS students able to understand this type of humanistic material?''
I found that the answer was broadly yes. Students were able to have cogent discussions about it and apply it appropriately.
The works they submitted reflected a coherent understanding of the content, as exemplified in the works shown in \autoref{fig:enter-label}.
It was inspiring to see students from diverse cultural and academic backgrounds encountering the situated and non-neutral nature of data for the first time and grow into being more critical thinkers.

While some students noted that this was the first class to make them apply ideas in this way, I suggest it need not be the last.
A reviewer of this work wondered ``How is this useful for CS students?'' I think that finding places to engage in modes of deep, careful, critical thinking that are outside the normal regime of ethics (as guided by ACM curricula~\cite{acm}) is valuable.
Ethical thinking is useful, but critical thinking---as manifest through critical theory---offers an intriguing alternative approach to developing key skills.
Moreover, it provides useful analytic skills and helps build an awareness of the inherently sociotechnical nature of even the most technical domains.

\parahead{Content restriction}
A complication (perhaps self-imposed) was my own anxieties around conducting this class in the specific moment in time. What content would be ``allowed''? What would negatively affect my tenure case? For instance, while not specifically forbidden by Utah law, it was strongly suggested to me that I not teach critical race theory as part of this course (despite its relevance), and so I did not. While there are assurances of academic freedom, addressing the tug-of-war of institutional acceptability with topical importance felt bracing to parse---particular as a new professor. As others explore similar courses, developing a shared curriculum may usefully provide some shared standardization through institutionalization.

\parahead{Interconnections}
A key accidental success was how the various works talked to each other.
While typical bibliographic conversations took place (for instance, Akbaba et. al's~\cite{akbaba2024entanglements} work on entanglement directly connects to data feminism), another type of recurrence was how works might rhyme or involve a repeated character's appearance. For instance, Wittgenstein appeared a number of times; sometimes as an obstinate caricature of analytic philosophy,
and at others as highlighting the inherent limitations of language.
Similarly, Latour was often referenced (to various effects but often as a synecdoche for Science Technology Studies) before we actually got to reading one of his works---giving a monster-at-the-end-of-the-book style build up.
In a more traditional course, the conceptual scaffolding used to help students situate and connect topics might be provided by reading works chronologically or by building to punchlines (like spending weeks on mechanics to get the grandeur of general relativity), but the multi-threaded nature of the topic does not align with such structures. So instead, I found that such characters kept things feeling familiar and connected in  ways that felt grounding and useful.
Moreover, I felt like having these characters made these ideas feel more concrete and human, even if the particular versions of these people were reductions of the full scope of their works.

\subsection{The Scavenger Hunt}

The other key element of this content was its form.
I used the scavenger hunt structure partly as an antidote to a particular form of course I encountered reasonably often in my own education, in which students read a bunch of papers and then do a bad project not relevant to their interests.
While there is extrinsic value in this form of course (learning to read a lot of papers is, in fact, wonderful), there is relatively little direct value. In adopting the scavenger hunt format, I attempted to address this issue by allowing students to just do things that were valuable to them (and, ideally, help them practice some time management skills).

While I stand by this goal, at the two-thirds mark, no one had turned anything in. So as to not fail the entire class, I held one-on-ones with students to help them make plans to pass the class.
While students mostly kept to these self-imposed deadlines, it was an unfortunate outcome that submission styles ended up being so traditional and that the corresponding pieces of work often ended up being rushed.
Based on an earlier experiment~\cite{mcnutt20capp}, I had believed that engagement with this structure would be more lively and need less structuring; so it was disappointing to see it turn out this way.

One complication was that the freeform structure of the course meant that class activities synchronized with the readings (as incentivized for engagement by being valued with points) were not possible. There were some topics that I think it would have been worthwhile to assign a problem set about, so that students could have worked through those topics in more concrete ways. For instance, while we spent significant time discussing Gregor's notions of theory~\cite{gregor2006nature} in class (and frequently drew on it as a means of theoretical analysis), it would have likely been better synthesized by having students deeply connect with a set of predefined theories.
Balancing structured academic engagement with free-form student benefit is a key trade-off of this design, meaning that it will work in some contexts but not others.

While these difficulties may be intrinsic to this course design, these snarls may have also been due to particular factors of my teaching, me, or my students.
I still believe in this method and think it is worth exploring more deeply.
For instance, one improvement that students brought up in post-course discussions would be to enforce mid-way deadlines (e.g., half the available points are due by spring break).
These and issues like how to scale the course to a less intimate number of students are worth exploring.

\section{Conclusion}

Finally, I want to briefly touch on the future. As the world enters an age that might be less receptive to considerations of power and humanistic value, I suggest that looking at those topics from within the recluse of computer science may well be useful.
Many non-CS departments seem to be scrambling to find ways to integrate engineering generally (or AI specifically) into their work. Finding ways to work humanistically from within CS may offer a useful way to meet those efforts and find areas of mutual resilience.

More specifically, as critical visualization pedagogy practice develops, it may be useful for those who have taught in this area to synthesize around a common curriculum. This would allow a more straightforward understanding of the contents of courses (and communication of what students know). Moreover, I do not think that my approach to understanding critical theory is the only one. I've heard others have been teaching similar courses (or starting to think about doing so). I call for us to continue to build and develop critical theories and practices in community.

\vfill
\pagebreak

\acknowledgments{
  This work was performed at Kahlert School of Computing, without whose support for this experimental class format, this work would not have been possible.
  Thanks as well to Georgia Panagiotidou and Derya Akbaba for providing comments both on the course and a draft of this report.
}

\bibliographystyle{abbrv-doi}
\bibliography{template}

\end{document}